# Magnon Hall effect and anisotropic thermal transport in NiFe and YIG ferromagnets

B.Madon[1], Do Ch. Pham[1], D. Lacour[2], A. Anane[3], R. Bernard[3], V. Cros[3], M. Hehn[2] and J.-E. Wegrowe[*1]

[1]Ecole Polytechnique, LSI, CNRS and CEA/DSM/IRAMIS, Palaiseau , France.
[2]Institut Jean Lamour UMR 7198 CNRS, Université de Lorraine, Vandoeuvre les Nancy France.
[3]Unité Mixte de Physique CNRS/Thales and Université Paris Sud, Palaiseau, France.
*Correspondence to:jean-eric.wegrowe@polytechnique.edu.

**Abstract:**

The Righi-Leduc effect refers to the thermal analogue of the Hall effect, for which the electric current is replaced by the heat current and the electric field by the temperature gradient. In both cases, the magnetic field generates a transverse force that deviates the carriers (electron, phonon, magnon) in the direction perpendicular to the current. In a ferromagnet, the magnetization plays the role of the magnetic field, and the corresponding effect is called anomalous Hall effect. Furthermore, a second transverse contribution due to the anisotropy, the planar Hall effect, is superimposed to the anomalous Hall effect. We report experimental evidence of the thermal counterpart of the Hall effects in ferromagnets, namely the magnon Hall effect (or equivalently the anomalous Righi-Leduc effect) and the planar Righi-Leduc effect, measured on ferromagnets that are either electrical conductor (NiFe) or insulator (YIG). The study shows the universal character of these new thermokinetic effects, related to the intrinsic chirality of the anisotropic ferromagnetic degrees of freedom.

**Introduction**

We report a new effect that could be added to the large family of thermokinetic transport phenomena. It consists in the observation of both anomalous Righi-Leduc effect - or magnon Hall effect [1,2]- and planar Righi-Leduc effect, measured on YIG and NiFe ferromagnets. The conventional *Righi-Leduc effect* is the thermal counterpart of the well-known *Hall effect*, and it accounts for the temperature gradient developed transversally to a heat current under a magnetic field. The adjectives *anomalous* and *planar* – that characterize the effect reported here - refer to the action of the *magnetization axial vector* (instead of a magnetic field) and the corresponding vector potential.

The application of a magnetic axial vector results in the partial breaking of two different symmetries. These symmetries are, on the one hand, the *invariance under time reversal* of the dynamical equations at the microscopic scale [3], and on the other hand, the *rotational invariance* (for an initially isotropic system). However, the symmetry breaking is *partial*. Indeed, in the first case, the time reversal invariance is recovered by the application of a $\pi-$rotation to the magnetization, and in the second case, the symmetry breaking is *partial* because the system is still invariant under any rotation around the magnetization. The consequence of these reduced symmetries is to impose a specific form to the heat transport coefficients [4] (see Supplementary), so that the temperature gradient becomes a very specific function of the magnetization states (as shown in Eq. (1) below).

Since the addition of a thin electrode in thermal contact with both edges of the ferromagnetic layer (see the set-up of Fig1) plays the role of a Seebeck thermometer (or thermocouple), the temperature difference ΔT is converted to a voltage difference ΔV, allowing the measurement of a magneto-voltaic signal. Such a device defines the principle of a magneto-thermal sensor. The studies of magneto-voltaic signals measured in response to thermal excitations on ferromagnetic layers has attracted considerable attention in the last years, with the observation of similar signals in conductor (NiFe), semiconductor (GaMnAs), and insulator (YIG) [5-13].

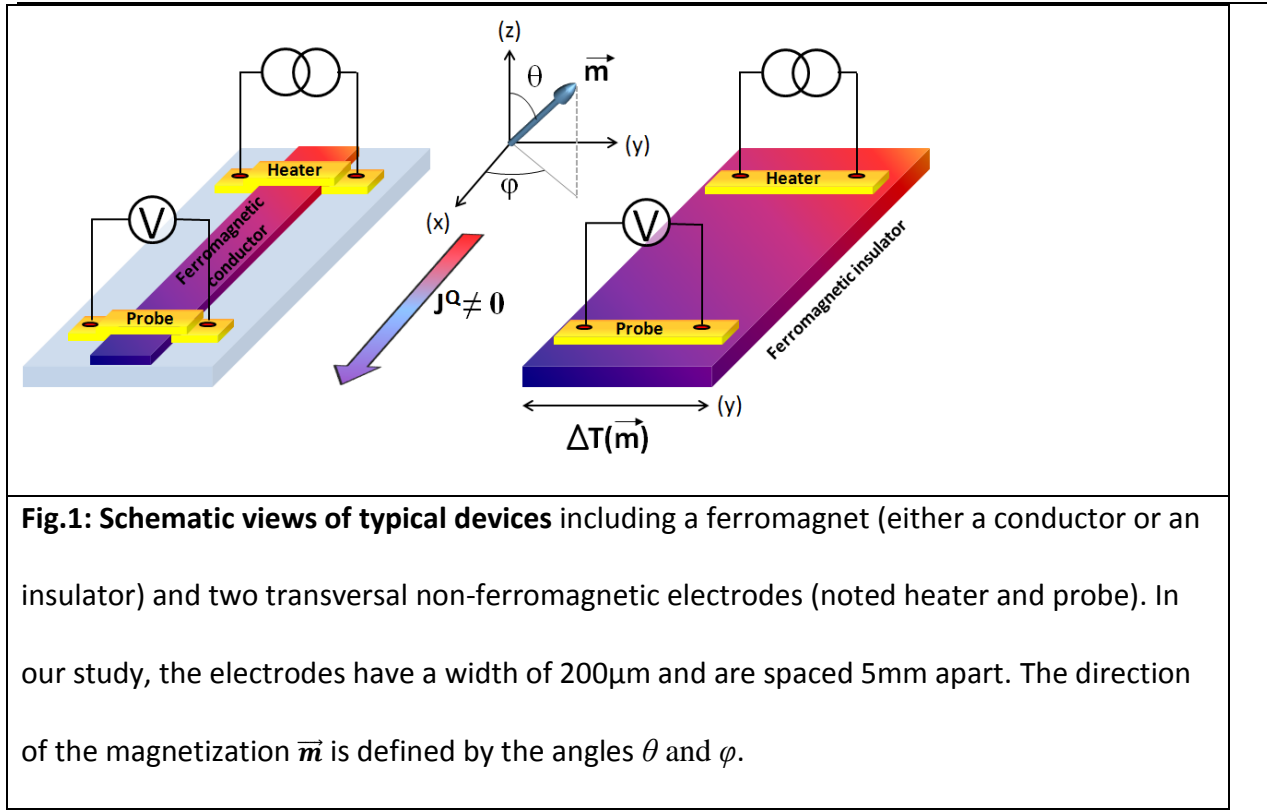

**Fig.1: Schematic views of typical devices** including a ferromagnet (either a conductor or an insulator) and two transversal non-ferromagnetic electrodes (noted heater and probe). In our study, the electrodes have a width of 200µm and are spaced 5mm apart. The direction of the magnetization $\vec{m}$ is defined by the angles $\theta$ and $\varphi$.

The anomalous Righi-Leduc effect has been predicted in various magnetic systems [14-17] and it has been measured recently in peculiar insulating ferromagnetic materials that possess a chiral crystalline structure [1,2]. The study of the anomalous Righi-Leduc effect in usual ferromagnetic layers (e.g. NiFe and YIG) has however been overlooked. On the other hand, the *planar Righi-Leduc effect* refers to the contribution of the anisotropy in the thermal conductivity. This anisotropy originates from the difference Δr between the thermal resistivity measured along the magnetization axis and the thermal resistivity perpendicular to the magnetization axis [18]. The comparative study between anomalous and planar Righi-Leduc effects, in NiFe and YIG ferromagnets, allows us to make a call in favor of a unifying interpretation in terms of anisotropic thermal transport (AThT), and to point out the universality of the phenomenon.

**Experimental angular dependences**

The samples contained electrodes that have been fabricated using the same set of shadow masks in a sputtering deposition system. They are fixed on top of two different magnetic materials. The first sample includes a 20nm thick $Ni_{80}Fe_{20}$ conductor stripe while the second sample contains a 20nm thick ferromagnetic YIG insulator [19]. Electrodes composed of Platinum (Pt) are deposited on top of each magnetic layer (see fig. 1). An ac electric current $I(t) = I_0 \cos(\omega t)$ is injected into the heater electrode (the power is of the order of a fraction of Watt and the frequency is a fraction of Hz). It produces a heat current $J^Q(t) = cRI_0^2(\cos(2\omega t)+1)/2$ having twice the frequency of the electric current ($c$ is a constant that takes into account the power dissipation (see Supplementary). The voltaic response $\Delta V_y$ to the thermal excitation is measured using a lock-in method *via* a probe electrode placed 5mm away from the heater electrode. All the measurements are done under a magnetic field $H$ of 1 Tesla that serves to rotate the magnetization. The voltage $\Delta V_y(\theta_H, \varphi_H)$ has been recorded for the two aforementioned devices either by varying the azimuthal angle $\varphi_H$ while keeping the polar angle $\theta_H$ fixed to 90° or by varying the polar angle keeping $\varphi_H$ fixed to 90° (see fig. 2).

First we observe that in both cases (conductor and insulator ferromagnetic material), $\pi$-periodic signals are measured in the magnetization in-plane (IP) configuration, while $2\pi$-periodic signals are observed in the out-of-plane (OOP) configuration. Second we find that the angular voltage variations display opposite phase on NiFe/Pt and on YIG/Pt. Finally, a triangular rather than a sinusoidal feature is observed in the NiFe sample for the measurements under an out-of-plane field.

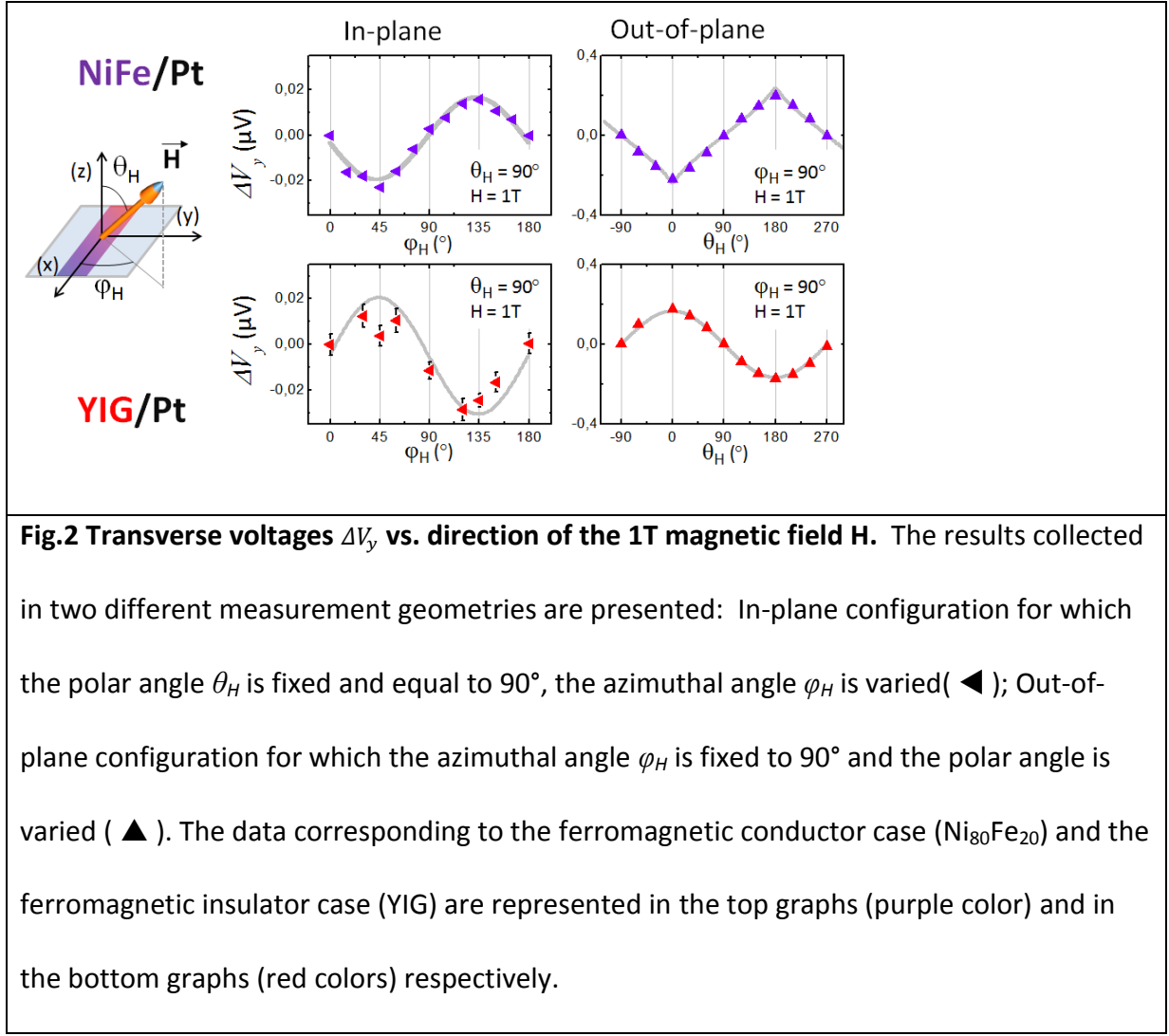

**Fig.2 Transverse voltages $\Delta V_y$ vs. direction of the 1T magnetic field H.** The results collected in two different measurement geometries are presented: In-plane configuration for which the polar angle $\theta_H$ is fixed and equal to 90°, the azimuthal angle $\varphi_H$ is varied (◄); Out-of-plane configuration for which the azimuthal angle $\varphi_H$ is fixed to 90° and the polar angle is varied (▲). The data corresponding to the ferromagnetic conductor case ($Ni_{80}Fe_{20}$) and the ferromagnetic insulator case (YIG) are represented in the top graphs (purple color) and in the bottom graphs (red colors) respectively.

**Anisotropic Thermal Transport (AThT)**

According to the AThT phenomenology (see Supplementary), a heat current $\vec{J}^Q$ injected inside the ferromagnet generates a thermal gradient $\vec{\nabla}T(\theta,\varphi)$ related to the orientation of the magnetization $(\theta,\varphi)$. Its description, based on the anisotropic Fourier equation, is valid both for the electric conductor and for the insulator. The probe electrode serves as a thermocouple that converts a local transverse temperature difference $\Delta T$ into a voltage $\Delta S. J_x^Q \Delta r$. The parameter $\Delta S$ stands for the difference between the Seebeck coefficients of the materials that compose the device. Considering that the heat current is along the *x* direction, the transverse voltage $\Delta V_y$ is given by the expression [4]:

$$\Delta V_y = \Delta S. J_x^Q \left( \frac{\Delta r}{2} \sin^2(\theta).\sin(2\varphi) + r_{RL} \cos\theta \right) \qquad \text{Eq.(1)},$$

where $\Delta r$ is the planar Righi-Leduc coefficient and $r_{ARL}$ is the anomalous Righi-Leduc coefficient.

From Eq.(1), the periods observed in Fig.2 can be easily understood. The 2φ term allows to explain the π-periodicity in the IP configuration while the 2π-periodic signals are linked the *cos*(θ) term that occur only in the OP configuration. Moreover we can also predict from Eq.(1), that the magnitude of the oscillations are equal to $\Delta S. J_x^Q \Delta r$ for the IP configurations and equal to $\Delta S. J_x^Q r_{ARL}$ for the OOP configuration. Using an independent measurement setup, we have determined ΔS for the NiFe and YIG based devices to be respectively -16.2µV.K$^{-1}$ and 0.69µV.K$^{-1}$ (see Supplementary). The opposite signs of ΔS provide a straightforward explanation for the aforementioned "antiphase" feature observed in Fig.2 comparing the results on NiFe/Pt and YIG/Pt. Finally taking into account the magnetic properties of the ferromagnetic layers (see Supplementary)**,** we were able to fit all measurements the only free parameters were either $\Delta S. J_x^Q \Delta r$ (in the IP configuration) or $\Delta S. J_x^Q r_{ARL}$ (in the OP configuration). It can be seen on Fig.2 (grey lines) that all the experimental results are in excellent agreements with our interpretation based on anisotropic thermal transport in the ferromagnet. The triangular profile (rather than sinusoidal) exhibited by the $Ni_{80}Fe_{20}$ device in the OP configuration is simply due to the fact that a 1 Tesla magnetic field is not large enough to fully saturate the magnetization perpendicular to the plane of the film (see Supplementary).

To test the robustness of the AThT explanation, we have first varied the thickness of the Pt probe from 5nm to 100nm. Defining the maximum amplitude of the magneto-voltaic signal $\delta(\Delta V_y) = \Delta V_y(180°) - \Delta V_y(0°)$. A decrease of $\delta(\Delta V_y)(d)$ as a function of the probe thickness is observed (Fig.3a). Such a decrease is often interpreted as the effect of spin injection at the interface [20-22]. Here we demonstrate that the thermal shunt effect suffices to explain the data. Not all the injected heat current $J_x^{*Q}$ contributes to the AThT effect since a part of this current is also flowing into the neutral Pt electrode. In order to evaluate the active part of the heat current, we can rewrite:

$$J_x^Q = \frac{\rho_{Th}^{Pt} . d_{NiFe}}{\rho_{Th}^{Pt} . d_{NiFe} + \rho_{Th}^{NiFe} . d} J_x^{*Q} \qquad \text{Eq.(2),}$$

assuming a simple scheme of two thermal conductors in parallel, as for anisotropic Hall measurements [23] (see Supplementary). The dependence of the signal on the thickness d of the Pt probe is calculated using the tabulated values $1/\rho_{Th}^{Py} = 72$ W.m$^{-1}$.K$^{-1}$ for Ni$_{80}$Fe$_{20}$ and $1/\rho_{Th}^{Pt} = 46$ W.m$^{-1}$.K$^{-1}$ for Pt, and the value of $\Delta S . J_x^{*Q} . r_{ARL}$ presented in Fig. 2 (see Supplementary). From the good agreement between the experimental curve and the prediction of Eq.(2) in Fig.3a), we conclude that the sole thermal shunt effect suffices to reproduce the observed decrease (the electrical counter part of the shunt effect is also reproduced without adjustable parameter, as shown in of Supplementary Fig. S9.

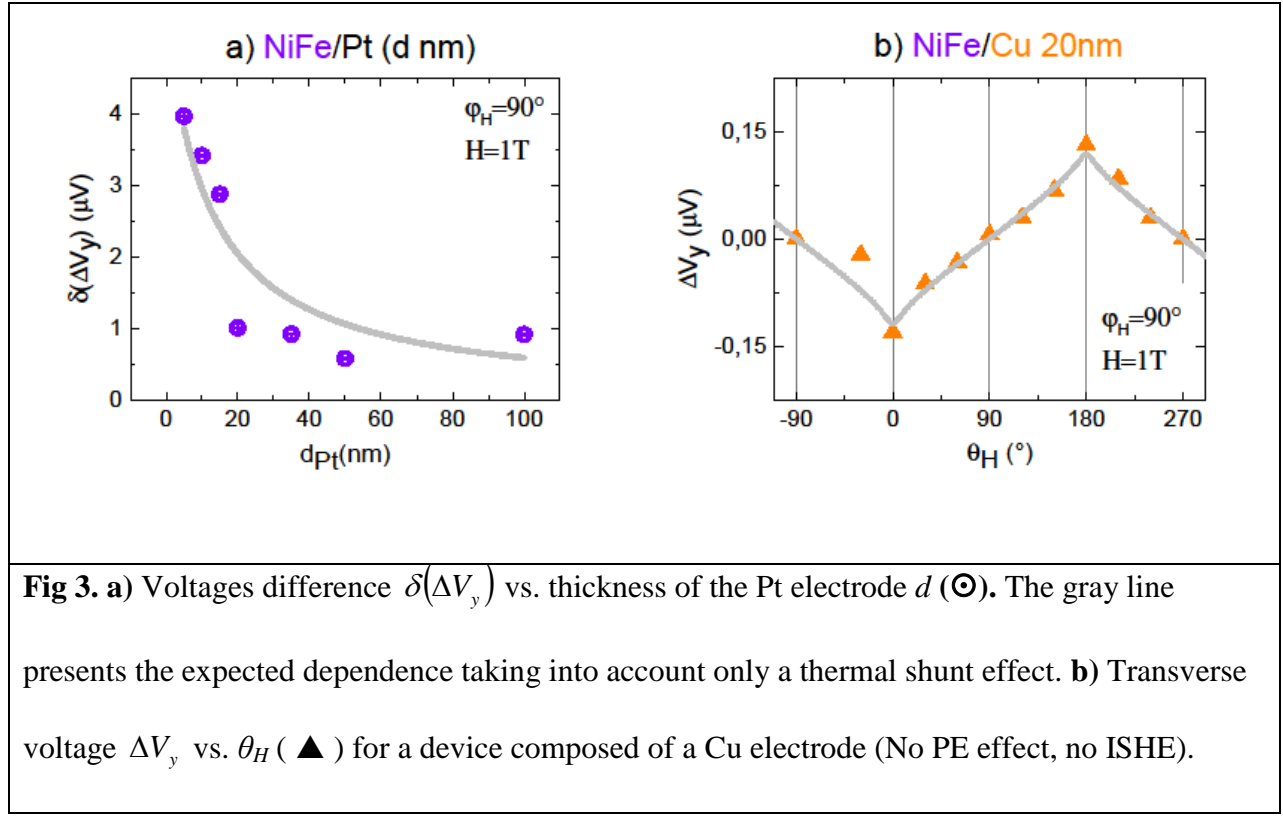

**Fig 3. a)** Voltages difference $\delta(\Delta V_y)$ vs. thickness of the Pt electrode $d$ (⊙). The gray line presents the expected dependence taking into account only a thermal shunt effect. **b)** Transverse voltage $\Delta V_y$ vs. $\theta_H$ (▲) for a device composed of a Cu electrode (No PE effect, no ISHE).

Moreover, the AThT interpretation does not require to invoke the hypothesis on inverse spin Hall effect (ISHE) or of a proximity effect arising from induced magnetic moments [11, 13]. In order to verify this statement, we have replaced the Pt electrodes by ultra-pure copper ones (99.9999% purity target). Indeed, the use of Cu electrodes allows to test at the same time the ISHE and PE hypotheses since both effects are absent in Cu [24,25,11]. We observed the magneto-voltaic signal even with pure copper electrode (Fig .3 b), as expected for AthT.

**Discussion and Conclusion.**

We have observed the coexistence of both anomalous and planar Righi-Leduc contributions in NiFe and YIG, of comparable amplitudes (leading to a transverse temperature difference of the order of 10 mK).

Although the *anomalous Righi-Leduc* effect can simply be understood on the basis of the Onsager reciprocity relations [3] (Supplementary Equations S2), a microscopic description can also been performed with a dedicated *vector potential* – or the corresponding *local gauge* and *Berry phase* – associated to the ferromagnetic system under consideration [26-28,14-17]. This problem generalizes sixty years of intensive theoretical development related to the *anomalous Hall effect* (starting with the work of Karplus and Luttinger in 1954 [29], and summarized e.g. in the review by Nagaosa et al [30]). Like the Lorentz force in the case of the conventional Hall effect, and like the spin-orbit scattering force in the case of anomalous Hall effect, the transversal force measured in this study can be derived from a vector potential. This force is thus neither conservative (it cannot be derived from a scalar potential) nor dissipative (no power can be extracted).

A second transverse force is observed, which is generated by the anisotropy of the ferromagnetic excitations $\Delta r \neq 0$. The measurements show that the two forces are not independent: the *anomalous Righi-Leduc* coefficient is associated to the *planar Righi-Leduc coefficient*. The same ferromagnetic axial vector is indeed responsible for both the anisotropy of the heat resistance ($\Delta r \neq 0$) and the breaking of the time invariance symmetry.

In conclusion, our results show that the anomalous Righi-Leduc effect, which has already been observed in specific ferromagnetic structures, is universal. This effect is observed in parallel to the planar Righi-Leduc effect. Both planar and anomalous Righi-Leduc effects should be present in any ferromagnetic materials in the same manner as anomalous and planar Hall effects can be expected *a priori* in any ferromagnetic conductors.

**Acknowledgements**

The Authors acknowledge H. Molpeceres and A. Jacquet for their assistance and O. d'Allivy Kelly for fruitful discussion. Financial funding RTRA `Triangle de la physique' Projects DEFIT n° 2009-075T and DECELER 2011-085T, the FEDER, France, La région Lorraine, Le grand Nancy, ICEEL and the ANR-12-ASTR-0023 Trinidad is greatly acknowleged.

# Supplementary Materials

I **Magnetic and electric characterization of the 20 nm thick permalloy (Ni$_{80}$Fe$_{20}$) samples.**

**I -1. Ferromagnetic quasi-static states.**

Due to the thin layer structure, the magnetization of the Permalloy (**Ni$_{80}$Fe$_{20}$ or** Py) layer is single domain. As a consequence, the magnetization $\vec{M} = M_s \vec{m}$ is a vector of constant modulus $M_s$ (magnetization at saturation) oriented along the unit vector $\vec{m}$. The quasi-static magnetization states are given by the minimum of the ferromagnetic free energy. This energy depends on three parameters, namely the magnetization at saturation $M_s$, the demagnetizing field $H_d$, and the magnetocrystalline anisotropy field $H_{an}$, confined in the plane of the layer. The corresponding energy is the sum of the three terms:

$$F = -\vec{H}.\vec{M} + \frac{1}{2} H_{an} M_s \sin^2 \xi_a + \frac{1}{2} H_d M_s \cos^2 \theta \qquad \text{Eq.(S1)}$$

where $\xi_a = (\vec{H}_{an}, \vec{m})$ is the angle between the magnetocrystalline anisotropy axis and the magnetization, and $\theta$ is the angle between the vector $\vec{n}$ normal to the plane of the layer and the magnetization.

The minimum of the energy $F$ (**Eq.(S1)**) sets the position of the magnetization, i.e. the radial angle $\theta$ and the azimuthal angle $\varphi$ as a function of the amplitude $H$ and direction $\theta_H$ and $\varphi_H$ of the applied field. The minimum is calculated through numerical methods (Mathematica® program).

The magnetization states were characterized using anisotropic electric transport properties, with the use of three different experimental configurations, which correspond to anisotropic magnetoresistance (AMR) **(31)**, planar Hall effect (PHE), and anomalous Hall effect (AHE)**(30)**.

**I-2. Electric properties**

The electric transport is described by the Ohm's law that relates the electric field $\vec{\varepsilon}$ to the electric current $\vec{J}^e$ with the use of the conductivity tensor: $\vec{\varepsilon} = \hat{\rho}.\vec{J}^e$ (note that for convenience, the experiments are usually performed in a galvanostatic mode, i.e. with constant current distribution $\vec{J}^e$). For a polycrystalline conducting ferromagnet, the conductivity tensor $\hat{\rho}$ is defined by three parameters. If the reference frame is such that the unit vector $\vec{m}$ is aligned along Oz, the parameters are the resistivity $\rho$ measured perpendicular to magnetization, the resistivity $\rho_z$ measured parallel to magnetization, and the Hall cross-coefficient $\rho_H$. According to Onsager reciprocity relation $\rho_H = \rho_{xy} = -\rho_{yx}$ and we have in the reference frame $\{x,y,z\}$:

$$\hat{\rho} = \begin{pmatrix} \rho & \rho_H & 0 \\ -\rho_H & \rho & 0 \\ 0 & 0 & \rho_z \end{pmatrix}$$

Accordingly, the Ohm's law can be expressed in an arbitrary reference frame, as (31):

$$\vec{\varepsilon} = \rho \cdot \vec{J}^e + (\rho_z - \rho)(\vec{J}^e \cdot \vec{m})\vec{m} + \rho_H \vec{m} \times \vec{J}^e$$

or, explicitly:

$$\vec{\varepsilon}(\theta, \varphi) = \begin{pmatrix} (\rho + \Delta\rho m_x^2)J_x^e + (\Delta\rho m_x m_y - \rho_H m_z)J_y^e + (\Delta\rho m_x m_z + \rho_H m_y)J_z^e \\ (\Delta\rho m_x m_y + \rho_H m_z)J_x^e + (\rho + \Delta\rho m_y^2)J_y^e + (\Delta\rho m_y m_z - \rho_H m_x)J_z^e \\ (\Delta\rho m_x m_z - \rho_H m_y)J_x^e + (\Delta\rho m_y m_z + \rho_H m_x)J_y^e + (\rho + \Delta\rho m_z^2)J_z^e \end{pmatrix} \quad \textbf{Eq. (S2)}$$

where $\Delta\rho = \rho_z - \rho$, $m_x = \sin\theta\cos\varphi$, $m_y = \sin\theta\sin\varphi$, $m_z = \cos\theta$. The angle $\theta$ is the same radial angle as the one introduced in the magnetic free energy, $\varphi$ is the azimuthal angle between the direction $Ox$ and the projection of the magnetization in the film plane. After integration, Eq. (S2) gives the *magneto-voltaic signals* that corresponds to the Anisotropic magnetoresistance (diagonal terms), the anomalous magnetoresistance (second term of the non-diagonal matrix elements), and the planar magnetoresistance (first term of the non-diagonal matrix elements). The same line of reasoning is applied in section **II-1** below for the transport of heat.

### I-2-1. Anisotropic magnetoresistance (AMR)

For AMR measurements, the voltage is measured along the same axis as the current flow (see Fig.1). The voltage is given by the integration over x of the first line in Eq. (2) with $J_z^e = J_y^e = 0$.

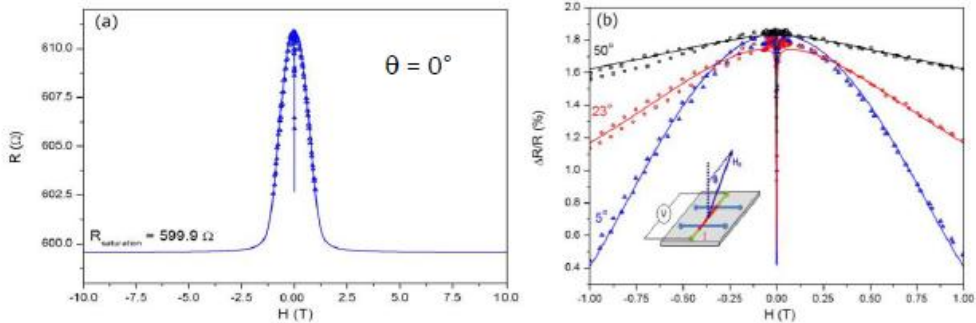

***Fig.S1***: *Resistance as a function of the amplitude of the external perpendicular field at $\varphi = 0°$ for (a) $\theta=0$ and (b) zoom for $\theta=5°$, $\theta=23°$ and $\theta=50°$. The points are the measured data and the line is the fit calculated from the minimization of the energy Eq.(S1) and* Eq.(S2).

Figure S1 shows the resistance as a function of the external perpendicular field at $\varphi = 0$. The fitted parameters are $H_d = 1T$ and the AMR ratio is found to be $\Delta R/R = 1.83\%$. Note that the saturation is not reached for $H=1T$. Consequently, the direction of the magnetization $(\theta,\varphi)$ does not exactly coincide with that of the external field $(\theta_H, \varphi_H)$: Indeed we have exploited this behavior in order to show that the magneto-voltaic signal is not a response to the external magnetic field (i.e. it is not the usual Nernst or Righi-Leduc effect), but a response to the magnetization (i.e. it is either the anisotropic Nernst or the anisotropic Righi-Leduc effect).

On the other hand, the in-plane magnetocrystalline anisotropy field $H_{an}$ is very weak, about $5.10^{-4}$ T, but its effect is rather dramatic as shown in Fig.2. In the vicinity of $\theta_H = 0°$ (modulo 180°), the magnetization suddenly switches from its initial position imposed by the applied field from $\varphi_H = 0°$ or $\varphi = 90°$ to $\varphi = 30°$ which is the direction of in plane anisotropy. This jump is well reproduced by the numerical simulation shown in Fig.S2.

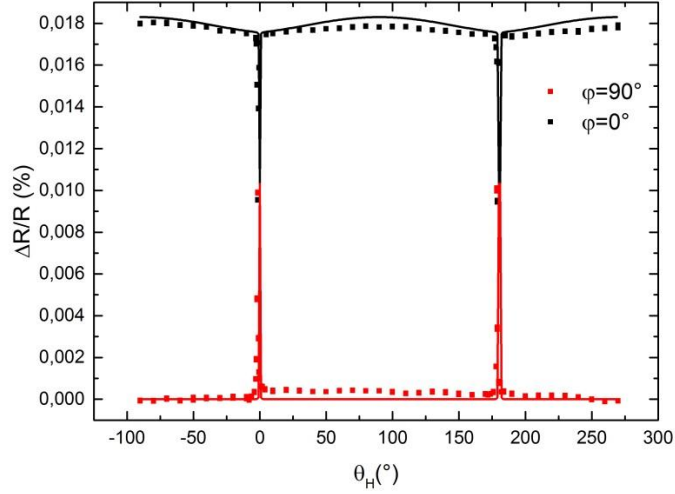

***Fig.S2***: *Magnetoresistance ratio (AMR) as a function of the out-of-plane angle $\theta_H$ for an external field of H=0.2T at $\varphi_H = \varphi = 0°$ (black upper curve), and at $\varphi_H = \varphi = 90°$ (lower curve). If $\theta$ is close to zero modulo 180°, the magnetization switches to the direction $\varphi \approx 30°$ (which corresponds to the plane defined by the external field and the anisotropy field).*

### I-2-2. Anomalous Hall effect (AHE) and Planar Hall effect (PHE)

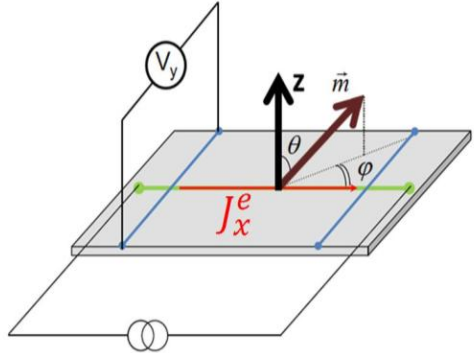

***Fig.S3***: *Configuration for AHE and PHE measurements.*

For planar Hall effect (PHE) and anomalous Hall effect (AHE), the electric current is injected along *0x* axis, but the voltage is now measured on the transverse electrode, along 0y (see Fig.S3). The voltage is given by the integration along the electrode of the second line of equation (2) with $J_z^e = J_y^e = 0$ :

$$V_y = I_x \left( \frac{A'}{2} \frac{\Delta R}{R} \sin^2\theta \sin 2\varphi + B\rho_H \cos\theta \right) \qquad \textbf{Eq. S3}$$

The first term is due to PHE while the second term is due to AHE. The coefficients *A'* and *B* are fitting parameters of the order of *L/A* where L is the distance between the two contacts and A is the section of the electrode (*A'* and *B* also include the contact resistance, so that they differ slightly from one sample to the other). The two contributions co-exist for an arbitrary direction of the magnetization, except if the configurations are fixed for the external magnetic field $\theta_H = 90°$ (in plane measurements as a function of $\varphi$ for pure planar Hall effect) or at $\varphi_H=0°$ or $\varphi_H=90°$ (out-of plane measurements as a function of $\theta$ for pure anomalous Hall effect).

Figure S4 shows out-of-plane measurements (AHE) as a function of the angle $\theta_H$, performed at (A) *H=0.2T* and *H=1T*. The calculated curve (continuous lines) follows closely the experimental data for *H=0.2T*. The jump of the magnetization for $\theta_H$ close to zero [resp. 180°] is that described on the AMR measurements presented in Fig.S2. The deviation between calculation and experimental data in Fig. S4(B) is explained by the metastable states due to the irreversible jump (the hysteresis loop is time dependent), that are not taken into account in the calculation of the quasi-static states.

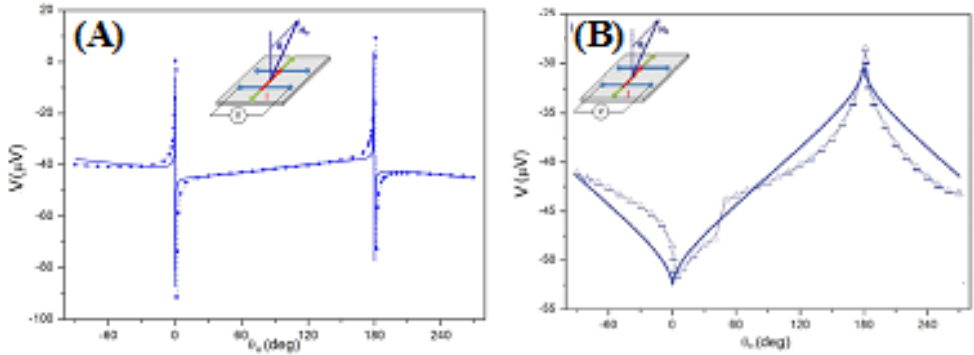

***Fig S4***: *(A) Out-of-plane (AHE) voltage as a function the angle $\theta_H$ for H=0.2T at $\varphi_H=0°$. (B) Same configuration for H=1T. The symbols are the experimental data and the line is calculated based on Eq.(S3) and on minimization of Eq.(S1).*

Figure S5 shows the in plane measurements with a saturation field of *H=1T*. The curve follows exactly the expected $\sin 2\varphi$ with a single adjustable parameter $R_H$.

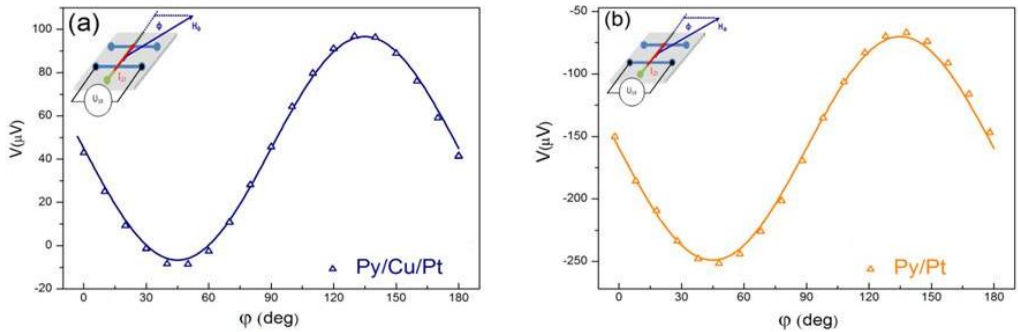

**Fig S5**: *Planar Hall voltage as a function of the angle $\varphi_H$ for an in-plane field ($\theta=\theta_H=90°$ of H=1T for the Cu and Pt electrodes. (a) Py(20nm)/Cu(5nm)/Pt(10nm) and (b) Py(20nm)/Pt(10nm). The presence of Cu does not change the magnetization states.*

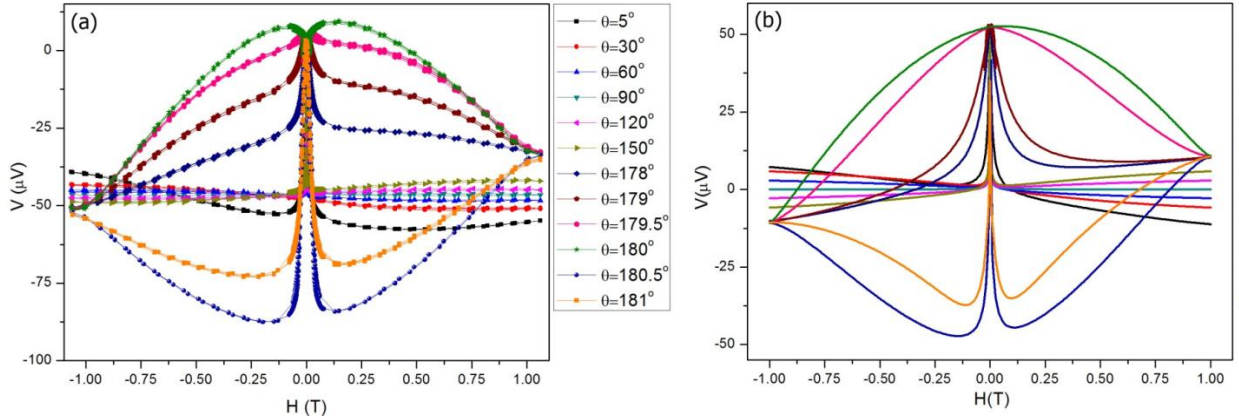

***Fig.S6***: *(a) Measurements of the Hall voltage as a function of the out-of-plane external magnetic field ($\varphi=0$) for different angle $\theta_H$. (b) Calculation based on Eqn.S1 and Eqn.S3 Planar Hall effect dominates. Note the brutal reversal from $\theta_H=180°$ to $\theta_H=180.5°$. It is the same as the one shown in Fig.S2 and Fig.S4.*

The measurements presented in Fig.S6 show that the magnetization states are well characterized by the simulation based on Eqn.S1 and Eqn.S3, and using the parameters fitted as described previously (with in-plane and out-of-plane angular dependence).

### I-2-3. AHE and PHE as a function of the thickness of the electrodes

Figure S7 shows the dependence of both AHE (a) and PHE (b) as a function of electrodes thicknesses ranging from 5nm to 100nm under an applied field of *H=1T*. The profile of the curve is not changed by the variation of the thickness, which means that the magnetization states are not impacted by the electrode thickness. Fig.S7 shows that the amplitude of the signal changes dramatically between 5 and 50 nm.

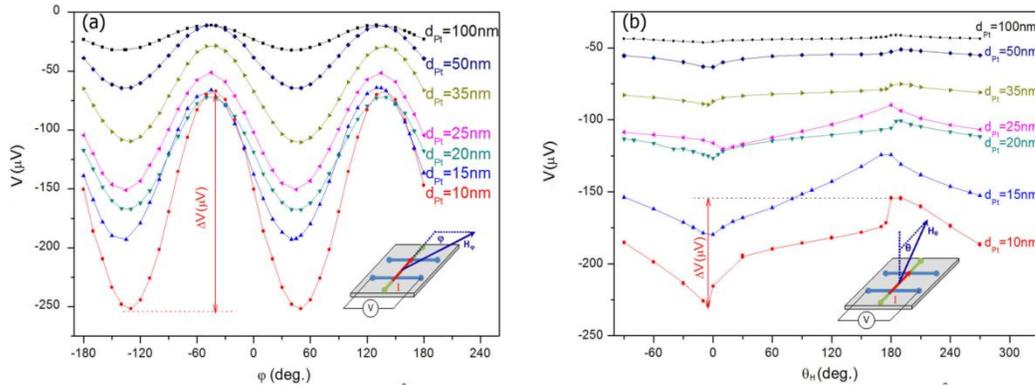

***Fig.S7***: *Measurement of the voltage for different thicknesses of the Pt electrode as a function of the angles at H=1T for (a) planar Hall effect ($\varphi_H = \varphi$) and (b) anomalous Hall effect ($\theta_H \neq \theta$). The signal $\Delta V$ is defined as the voltage difference between the maxima and minima.*

In order to justify the thickness dependence of the AHE and PHE signals, we first take the assumption that the non-ferromagnetic electrode is passive. The effective current that flows inside the ferromagnetic layer is not the initial current but it is divided into two branches (Fig. S8). A first branch is defined by the resistance of the ferromagnetic layer (shunt effect)(**23**).

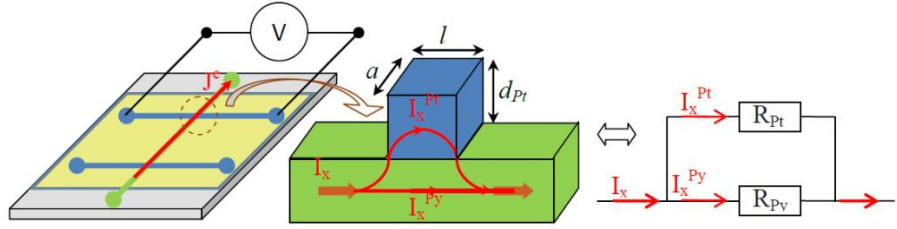

***Fig. S8***: *Illustration of the shunt effect that takes place at the level of the electrode. The effect is well described by a two resistor model $R_{Pt}$ and $R_{Py}$.*

The thickness dependence is given by the coefficient α such that $I_{eff} = α I$. We have:

$$\alpha = \frac{\rho_{Pt} d_{Py}}{\rho_{Pt} d_{Py} + \rho_{Py} d_{Pt}} \qquad \textbf{Eq.S4}$$

The Py thickness is $d_{Py}$ and that of the Pt electrode is $d_{Pt}$. The corresponding resistivities are $\rho_{Py}$ and $\rho_{Pt}$ that have been determined by independent resistance measurements.

|  | $\Delta V_y$; $\theta_H = 90°$ | α | $\frac{\Delta R}{R}$ (%); $\theta_H = 90°$ | $\Delta V_y$; $\varphi_H = 0°$ | $R_H$ |
|---|---|---|---|---|---|
| Py/Pt(5nm) | 231.3μV | 0.58 | 0.42 | 110μV | 0.095 |
| Py/Pt(10nm) | 186.8μV | 0.40 | 0.49 | 75μV | 0.094 |
| Py/Pt(15nm) | 130μV | 0.31 | 0.44 | 56μV | 0.090 |
| Py/Pt(20nm) | 98μV | 0.25 | 0.41 | 28μV | 0.056 |
| Py/Pt(25nm) | 98.5μV | 0.21 | 0.49 | 30μV | 0.071 |
| Py/Pt(35nm) | 82.6μV | 0.16 | 0.55 | 15μV | 0.047 |
| Py/Pt(50nm) | 53.7μV | 0.11 | 0.52 | 13.1μV | 0.060 |
| Py/Pt(100nm) | 19.7μV | 0.06 | 0.35 | 5.1μV | 0.043 |

***Table.S1***: *Parameters used for the calculation of Fig.S9*

The typical profiles of the thickness dependence of both the AHE and ANE are presented in Fig.S9. The measured data follows perfectly the profile predicted taking into account the shunt effect. There is no adjustable parameter in the calculation. We took the mean values of <ΔR/R> and <$R_H$> obtained by averaging the parameters (Table.S1) over all samples.

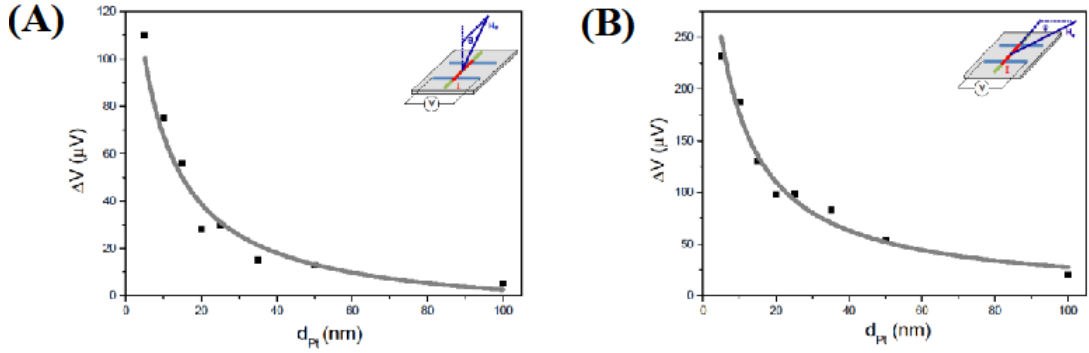

***Fig.S9***: *(A) Anomalous and (B) planar Hall signals ΔV as a function of the thickness d of the Py electrode. The points are the measured data and the line is the correction (coefficient α) due to the shunting effect Eq.(4).*

The excellent agreement between experiments and the predictions shown in Fig.S9 bring as a clear conclusion that the typical thickness dependence is only due to the shunting effect.

## II) Anisotropic Thermal Thransport (AThT)

### II-1 The anisotropic Fourier equation

In order to describe the transport of heat in a ferromagnetic system, we follow an equivalent approach of the one used in section I-2. Indeed, both electric and thermal transport phenomena obey the same symmetry properties, namely the rotational invariance of the system through any rotation around the magnetization axis and the time reversal invariance associated to the rotation $\pi$. The Fourier law takes thus the same form as the Ohm's law (the electric current is replaced by a heat current and the electric field by a gradient of temperature).

Fourier law relates the gradient of the temperature $\vec{\nabla}T$ to the electric current $\vec{\nabla}T = \hat{r}\vec{J}^Q$. The conductivity tensor $\hat{r}$ of a polycrystalline conducting ferromagnet (this is the case of the NiFe samples) is defined by three parameters. If the reference frame is such that the unit vector $\vec{m}$ is along $Oz$, we define the thermal resistance r measured perpendicular to the magnetization, the thermal resistance $r_z$ measured parallel to the magnetization, and the Righi-Leduc cross-coefficient $r_{ARL}$. According to Onsager reciprocity relation, we have in the reference frame $\{x,y,z\}$:

$$\hat{r} = \begin{pmatrix} r & r_{ARL} & 0 \\ -r_{ARL} & r & 0 \\ 0 & 0 & r_z \end{pmatrix}$$

The Fourier's law can then be expressed in an arbitrary reference frame, as **(4)**:

$$\vec{\nabla}T = r.\vec{J}^Q + (r_{//} - r)(\vec{J}^Q.\vec{m})\vec{m} + r_{ARL}\vec{m} \times \vec{J}^Q$$

where $\Delta r = r_z - r$. Explicitly:

$$\vec{\nabla}T(\theta,\varphi) = \begin{pmatrix} (r+\Delta r m_x^2)J_x^Q + (\Delta r m_x m_y - r_{ARL} m_z)J_y^Q + (\Delta r m_x m_z + r_{ARL} m_y)J_z^Q \\ (\Delta r m_x m_y + r_{ARL} m_z)J_x^Q + (r+\Delta r m_y^2)J_y^Q + (\Delta r m_y m_z - r_{ARL} m_x)J_z^Q \\ (\Delta r m_x m_z - r_{ARL} m_y)J_x^Q + (\Delta r m_y m_z + r_{ARL} m_x)J_y^Q + (\rho + \Delta r m_z^2)J_z^Q \end{pmatrix} \quad \textbf{Eq.(S5)}$$

where $m_x = \sin\theta\cos\varphi$, $m_y = \sin\theta\sin\varphi$, $m_z = \cos\theta$, $\theta$ is the same as the one introduced in the magnetic free energy and $\varphi$ is the angle between the direction $Ox$ and the projection of the magnetization in the plane of the sample.

The temperature difference $\Delta T_y$ can be measured between the two edges of the ferromagnetic layer along $0y$, thanks to the thermocouple effect. The voltage is given by the Seebeck coefficient $\Delta S$, such that $\Delta V_y = \Delta S \, \Delta T_y$ (see II-3 below). Since the heat current is mainly along $Ox$, we obtain the main equation used in this study:

$$\Delta V_y \approx J_x^Q \Delta S \left( \frac{\Delta r}{2} \sin^2\theta \sin 2\varphi + r_{ARL} \cos\theta \right) \quad \textbf{Eq.(S6)}$$

The second term in the right hand side of Eq.(S6) – proportional to $\cos\theta$ - defines the anomalous Righi-Leduc coefficient $r_{ARL}$, that can be measured directly with setting $\varphi=0$ or $\varphi=90°$ (out-of-plane measurements). On the other hand, the first term in the right hand side of Eqn.S6 – proportional to $\sin(2\varphi)$ (in-plane angle) – defines the planar Righi-Leduc coefficient $\Delta r$, that can be measured directly with setting $\theta=\pi/2$ (in-plane measurements).

## II-2 AThT on NiFe sample

In complement to the measurements on NiFe ferromagnet presented in the main text, complementary results obtained with a Cu(5nm)/Pt(10nm) electrode are shown in Fig.S10 and Fig.S11. We observe that the results are identical to that corresponding to the Pt(50nm) presented in Fig.2 of the main text (after correction due to the shunting effect). We can conclude that the Cu(5nm) electrode deposited between the ferromagnet and the Pt does not modify the signals significantly, in agreement with Eq.(S6). The angular dependences (radial and azimuthal) for H=1T are plotted in Fig.S10, with the numerical simulation, according to equation (S6). Fig.S10 and Fig.S11 display supplementary measurements with Cu electrodes.

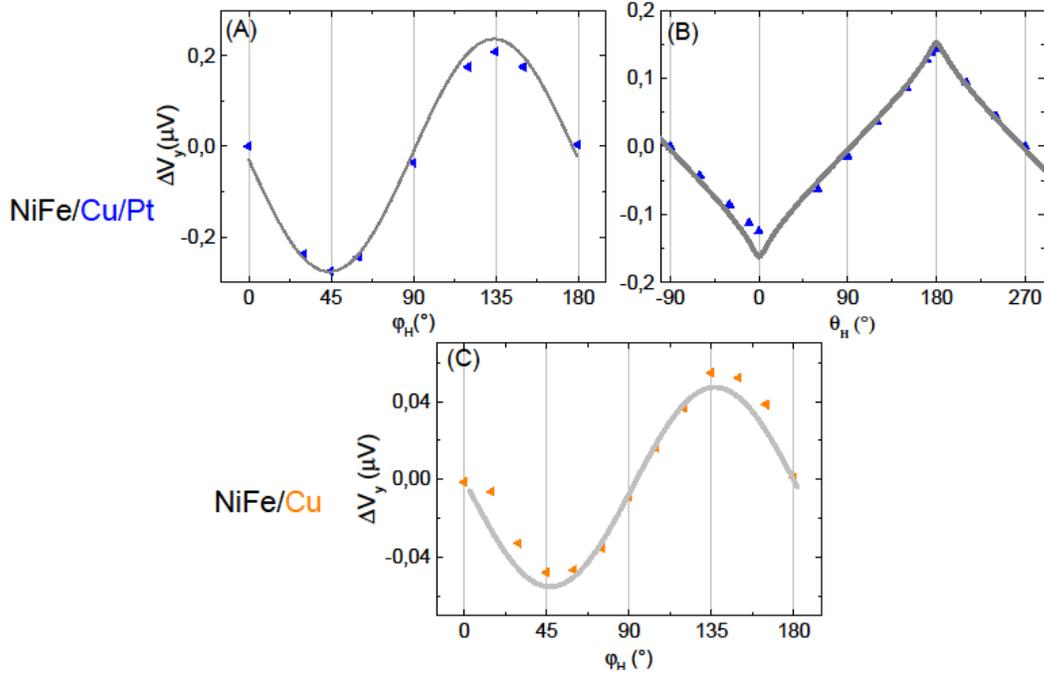

***Fig.S10*** *:(A-C) Transverse voltages vs. direction of the 1T magnetic field. Cu(5nm)/Pt(10nm) electrode: (A) In-plane configuration at $\theta_H = 90°$ and (B) out-of-plane configuration for $\varphi_H = 90°$. (C) Cu(20nm) electrode: in-plane configuration (see Fig3B for the out-of-plane configuration). The lines correspond to the calculation of Eq.(6) with minimization of the energy Eq.(1).*

The Anosotropic Thermal Transport (AThT) signals have been measured a function of the magnetic field (see Fig.11(a)) for three values of the direction of the applied field ($\theta_H$). The numerical simulations (continuous lines) are in excellent agreement with the experimental results. The magnetization reversal at small field is shown in the inset. The out-of plane angular variation *at $\varphi_H = 0°$* for a medium magnetic field (H = 0.18T) is plotted in Fig.S11(B). The irreversible jump of the magnetization (presented in Fig.S2 and Fig.S4) is clearly observed, and described by the numerical simulations.

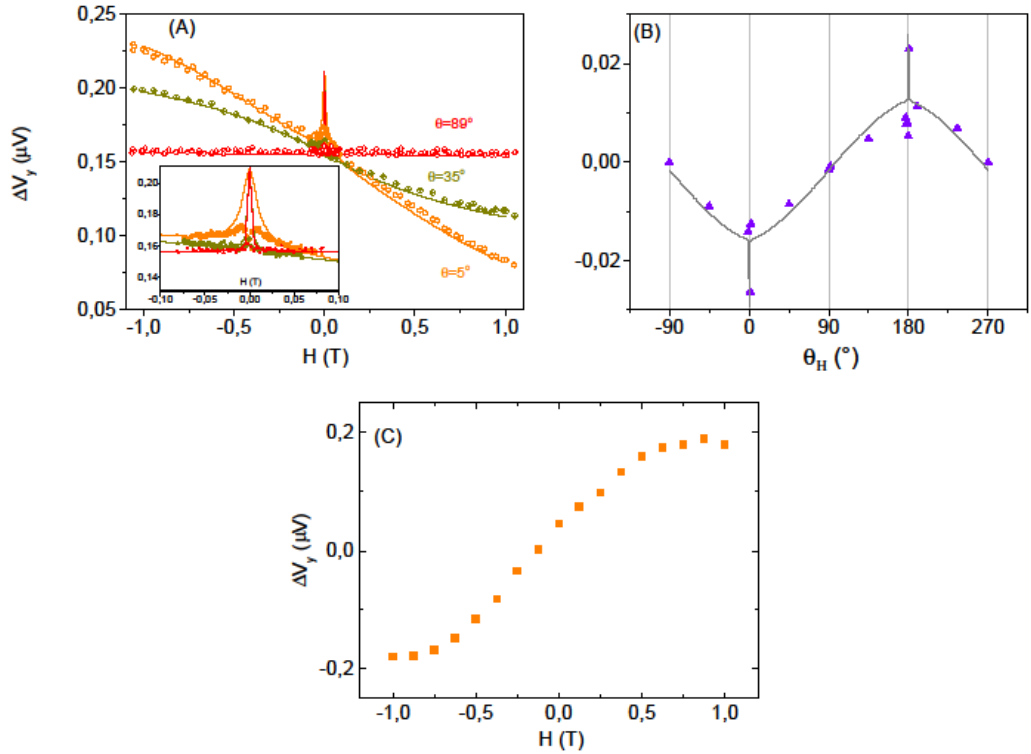

***Fig.S11***. *Transverse voltages on Py/Cu/Pt electrode as a function of (A) amplitude of the magnetic field H for three out-of-plane angles, (B) radial angle $\theta_H$ for an applied field of 0.18T. The line correspond to the calculation of Eq.(S6) with minimization of the energy Eq.(S1). (C). Transverse voltage on Py(40nm)/Cu(40nm) as a function of the amplitude of the magnetic field for out-of-plane configuration at 180°.*

## II-3. Heat power and magneto-voltaic signal

In our experiment, Joule heating is generated using AC current of pulsation $\omega$, injected into a resistance through a second electrode deposited on the ferromagnetic layer (see Fig.1 of the main text). The heat power flowing through the sample is proportional to the square of the current. As a consequence, the magneto-voltaic response to the heat excitation is measured at the double frequency $2\omega$.

We checked that the signal is proportional to the injected power as shown in Fig.12. The extrapolation to zero shows that the heat current $J^Q$ measured at the level of the electrode is simply proportional to the heat power injected by Joule effect: $J^Q_x = c\, P_{Joul}$, where the constant $c$ (such that $0<c<1$) contains all contribution of heat dissipation (including the coefficient $\alpha$). It depends on the frequency $\omega$ (see Fig.S13) and varies from one sample to the other. The change of the magneto-voltaic signal $\Delta U$ observed for different values of the out-of-plane external field (here for H=-1T and H=1T) is due to the anomalous Righi-Leduc (or anomalous magnon-Hall) effect studied in this work.

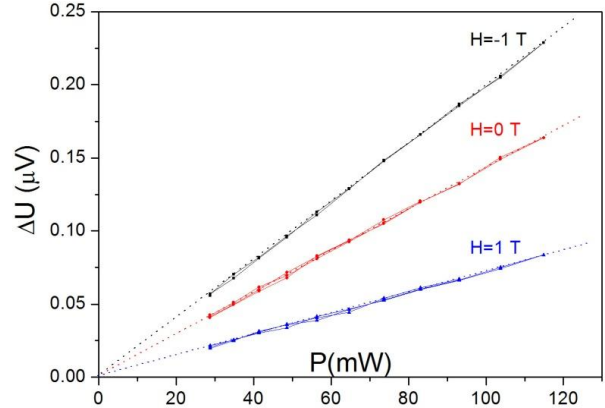

*Fig.S12: Measurement of the 2ω Magneto-voltaic signal as a function of the Joule power injected for different values of the out-of-plane applied field.*

The typical frequency we used is 0,01Hz. Smaller frequencies would lead to too long measurement time, while higher frequencies would give a too weak magneto-voltaic response. The amplitude $\Delta U(\omega)$ of the magneto-voltaic signal as a function of the frequency of the heat excitation is presented in Fig.13. This typical profile depends mainly on three characteristics, contained in the constant c, that are (i) the distance between the heat source and the electrode on which the magneto-voltaic signal is measured, (ii) the thermal conductivity of the substrate, and (iii) the electric contacts that thermally couple the sample to the voltmeters.

We checked, using a vacuum cell, that the heat dissipated through the surfaces of the layer does not affect the signal (see Fig.S13).

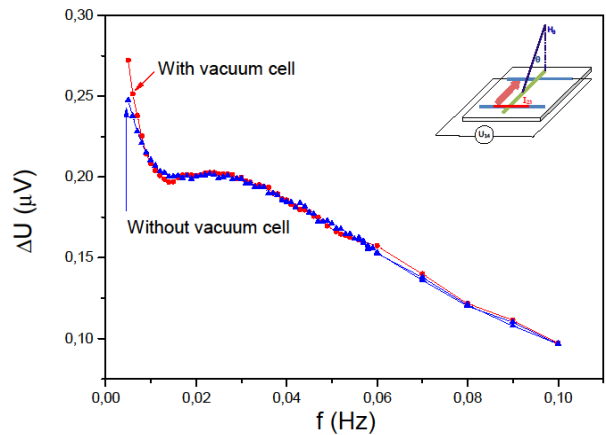

*Fig.S13. Frequency dependence of the 2ω magneto-voltaic signal. The typical profile is due to the thermal losses between the power injection and the measurement electrode. The local maximum at about 0.025 Hz is a good compromise between rapidity of the measure and amplitude of the signal. The curve has been measured in a vacuum cell (red points) in order to check that that dissipation throughout the surfaces is negligible.*

## II-4. Measurement of the thermocouples

In order to measure the Seebeck coefficient, we used a single line sample wired with aluminum wires and silver paint which is our reference material. One side of the sample was kept cool using an iced water bath and the other side was at room temperature. The voltage was measured over time. The maximum variation of $\Delta U(t)$ – that corresponds to the maximum temperature difference $\Delta T$ – gives a measure of the thermocouple $\Delta S = \Delta U / \Delta T$ (V/K). It is shown that the contribution of the Permalloy-Al interface is strong and negative (-16.2 µV/K) while the other contributions (Pt-Al, Pt-Ag, Pt-Au, etc…) are small and positive (< 1 µV/K). As a consequence, the thermocouple of NiFe dominates and the total thermocouple is always negative and of the order of -10 µV/K. In contrast, the thermocouple for Pt electrodes deposited on a YIG instead of NiFe (Pt/Al thermocouple), has a positive thermocouple of the order of +1 µV/K. *This sign inversion of $\Delta S$ explains the sign change observed between the magneto-voltaic signals of YIG ferromagnet and NiFe ferromagnet (see main text).*

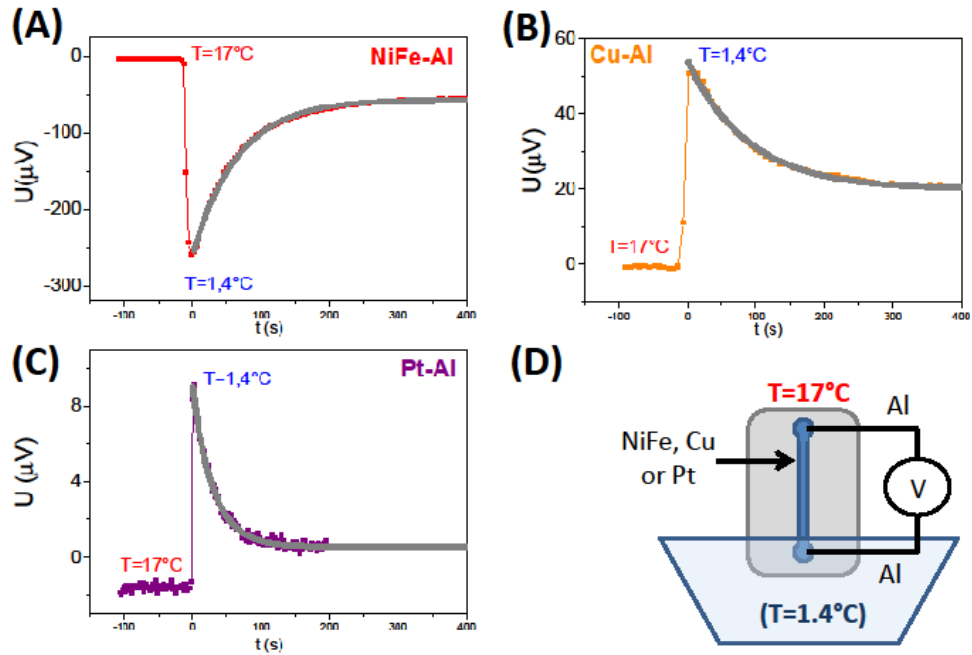

*Fig.S14. Time dependent measurement of the thermocouple generated by a contact wire of Al with the electrode of (A) NiFe (Py), (B) Cu, and (C) Pt. (D) sketch for the measure of the thermocouple. The temperature difference of $\Delta T = 15.6°C$ is imposed at t=0, and the relaxation due to thermalization of the metallic line is measured as a function of time. The calculated line is the exponential relaxation.*

The temperature difference measured between the two extremities of the electrode is typically $\Delta T = \Delta U / \Delta S = 0.002$ K. The transport coefficients related to anomalous Righi-Leduc effect (out-of-plane measurements) is found to be $c_1 r_{RL} = 0.16$ K/W for NiFe and $c_2 r_{RL} = 0.13$ K/W for YIG. The transport coefficient related to planar Righi-Leduc effect (in-plane measurements) is found to be $c_1 \Delta r = 0.07$ K/W for NiFe and $c_2 \Delta r = 0.02$ K/W for YIG. The unknown parameter $0.1 < c_i < 1$ (i={1,2}) takes into account heat dissipation between the heater and the electrode (including shunt effect) and varies from one sample to the other.